\documentclass[conference]{IEEEtran}
\IEEEoverridecommandlockouts

\usepackage{float}
\usepackage{listings}
\usepackage{hyperref}
\usepackage{multirow}
\usepackage{graphicx}
\usepackage{amsmath}
\usepackage[dvipsnames]{xcolor}
\usepackage{balance}
\usepackage{scalerel}
\usepackage{tikz}
\usetikzlibrary{svg.path}
\usepackage{caption}
\usepackage{academicons}
\usepackage{tabularray}
\usepackage{lipsum}
\usepackage{tcolorbox}
\usepackage{adjustbox}
\usepackage{mdframed}
\usepackage{orcidlink}
\usepackage{svg}
\usepackage{array}
\usepackage{ragged2e}
\usepackage{tabularx}
\usepackage{graphicx}
\usepackage[utf8]{inputenc}
\usepackage{colortbl}
\usepackage{longtable}
\usepackage{tikz}
\usepackage{amsmath}
\usepackage{dirtytalk}
\usepackage{algorithm}
\usepackage{algorithmic}
\usepackage{subcaption}

\def\BibTeX{{\rm B\kern-.05em{\sc i\kern-.025em b}\kern-.08em
    T\kern-.1667em\lower.7ex\hbox{E}\kern-.125emX}}
\begin{document}

\title{Cracking CodeWhisperer: Analyzing Developers' Interactions and Patterns During Programming Tasks}

\author{
Jeena Javahar\textsuperscript{1},
Tanya Budhrani\textsuperscript{1},
Manaal Basha\textsuperscript{1}, \\
Cleidson R. B. de Souza\textsuperscript{2},
Ivan Beschastnikh\textsuperscript{1},
Gema Rodríguez-Pérez\textsuperscript{1},\\
\textsuperscript{1}University of British Columbia, \quad
\textsuperscript{2}Federal University of Pará \\
\texttt{jeenajavahar@gmail.com},
\texttt{tanya.budhrani@connect.polyu.hk},
\texttt{manaals@student.ubc.ca},  \\
\texttt{cleidson.desouza@acm.org}, 
\texttt{bestchai@cs.ubc.ca}, \texttt{gema.rodriguezperez@ubc.ca}
}

\maketitle

\begin{abstract}
The use of AI code-generation tools is becoming increasingly common, making it important to understand how software developers are adopting these tools. In this study, we investigate how developers engage with Amazon's CodeWhisperer, an LLM-based code-generation tool. We conducted two user studies with two groups of 10 participants each, interacting with CodeWhisperer - the first to understand which interactions were critical to capture and the second to collect low-level interaction data using a custom telemetry plugin. Our mixed-methods analysis identified four behavioral patterns: 1) incremental code refinement, 2) explicit instruction using natural language comments, 3) baseline structuring with model suggestions, and 4) integrative use with external sources. We provide a comprehensive analysis of these patterns .
\end{abstract}



\section{Introduction}
Several IDE-based code generation tools have been released in the past few years, such as GitHub's Copilot~\cite{copilot}, Kite~\cite{kite}, Amazon's Code Whisperer~\cite{codeWhipserer}, Tabnine~\cite{tabnine}, and WPCode~\cite{WPcode}. 
Research reveals that being able to achieve their full potential requires a certain level of guidance to ensure that the tool's output aligns with the user's goal~\cite{Subramonyam2024}. 
%
%
Our research aims to explore \textbf{how developers engage with an LLM integrated into their IDE, and the patterns that emerge from their interactions across programming tasks}. 

To identify key interaction patterns, we conducted an initial study with 10 participants using Amazon CodeWhisperer in their IDEs to solve progressively challenging tasks. Screen recordings from this study provided insights into user behaviors and tool usage. Based on the initial findings from the first study, we developed CodeWatcher~\cite{basha2025codewatcher}, a tool that includes a VSCode plugin for capturing and analyzing developer interactions with LLM tools. We then conducted a second user study with 10 new participants using our tool. This provided us with low-level interaction data to study usage patterns. 

Our work aims to provide deeper insights into the practical usage of LLM-based tools for code generation by addressing the following research questions:

\begin{itemize}
    \item RQ1 - How do users interact with and edit code using CodeWhisperer to solve programming tasks, and what detailed interaction patterns emerge?         
    \item RQ2 - What fraction of CodeWhisperer's suggestions are retained by users?
\end{itemize}



As part of answering these research questions, we make the following contributions: 
\begin{enumerate}
    \item We bridge low-level interaction data with high-level LLM usage by recording and analyzing how users engage with LLMs 
    and how they edit code. This paper is one of the first studies to do so; and 
    \item We identify several new behavioural patterns associated with using LLM-based code generators.
\end{enumerate}

\section{Related work}
\label{sec:related}


LLMs are neural networks with billions of parameters pre-trained on large corpora of unlabeled text through supervised learning~\cite{overviewOfLLMsRESEARCHPAPER}. When applied to software development, LLMs can generate code from natural language descriptions.
In this paper, we use CodeWhisperer~\cite{codeWhipserer} now Amazon Q, a tool built by Amazon Web Services~\cite{CodeWhispererManual}. 

User studies of LLM-based code generators are broadly classified them into two groups: (i) studies of GitHub Copilot, and (ii) studies of ``customized'' code generation tools, i.e., tools created by the study authors as explained by Mendes et al.~\cite{Mendes2024}. Mendes et al. discussed two customized tools: NL2Code
~\cite{xu2022ide}, and GenLine
~\cite{jiang2022discovering}. Xu et al.~\cite{xu2022ide} concluded that most NL2Code's users had either a positive or neutral experience using it. Meanwhile, Jiang et al.~\cite{jiang2022discovering} discuss an evaluation of GenLine with 14 participants using it for a week: participants felt they had to ``learn'' how to interact with the tool because of unexpected model responses.

Barke et al. conducted a grounded theory analysis of GitHub's Copilot being used by 20 participants~\cite{groundedCopilot}, identifying two primary ways users interact with an LLM-based code generator: acceleration and exploration. Another study worth mentioning
is by Vaithilingam et al.~\cite{vaithilingam2022expectation} who conducted an experiment with 24 students using GitHub Copilot. Similar to Barke et al., students enjoyed using Copilot because it provided code that could be used as a ``starting point'', even when this code was incorrect. They also reported two coping strategies to deal with Copilot’s limitations: ``to accept the incorrect suggestion and attempt to repair it'' or simply stop using the tool.
Building on these previous studies, our work records users' interactions with CodeWhisperer and analyzes how users edit code. This approach bridges low-level interaction details with high-level LLM usage, offering an in depth view of developer engagement and the practical application of LLM-generated suggestions.


Contrary to some of the previous studies 
which rely on surveys, video recordings, interviews, etc~\cite{groundedCopilot, Mendes2024, Murali2024, dev-behav-eyetrack, vaithilingam2022expectation, trustInAIRESEARCHPAPER, xu2022ide} with a focus on user reactions,  
our paper uses low level interaction data to identify new behavioral patterns. 

To the best of our knowledge, only two papers adopt a similar approach to ours. Ziegler et al.~\cite{ziegler2022productivity}, analyzed internal Copilot developer usage data, including the completion shown to each developer, the completion accepted by the developer for inclusion in the source file, the number of characters in an accepted completion, among others. Tang et al.~\cite{tang2024codegrits}  who examined developer behaviour during validation and repair of LLM-generated code using eye-tracking and IDE actions with their tool, CodeGRITS. \textcolor{black}{By using the data captured by our tool-independent plugin~\cite{basha2025codewatcher}, we complement the previous studies and provide a new perspective to the study of interaction between developers and LLMs fine-tuned for code generation.}

\section{Methodology}
\begin{figure}[ht!]
    \centering
    \includegraphics[width=0.80\linewidth]{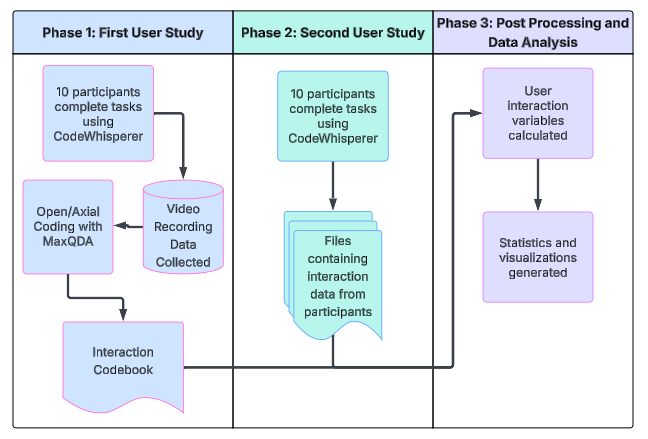}
    \caption{\textcolor{black}{Study Phases}}
    \label{fig:methodology}
\end{figure}

Our study was conducted in three phases (See Figure \ref{fig:methodology}). First, we carried out an initial user study using screen recordings to observe how participants interact with CodeWhisperer. This phase resulted in a detailed codebook outlining user interactions with CodeWhisperer. The codebook was used to create an IDE plugin, CodeWatcher~\cite{basha2025codewatcher}, that automatically logs users' interactions with the IDE by capturing significant changes in the text body and specific keystrokes. Second, we conducted a second user study, using the same setup as the study in phase 1 using CodeWatcher for recordings. Finally, we analyzed the logs to uncover patterns in user interactions. 

\subsection{Phase 1: Initial User Study}
\label{sec:phase1}
We recruited 10 undergraduate Computer Science students through posters, mailing lists, and student clubs. A screening survey collected demographics and programming proficiency, and we selected participants to ensure diversity in gender, native language, and experience. The study was approved by our university's ethics board and included five self-identified men and five self-identified women from varied linguistic and national backgrounds. None of the participants had prior experience with Amazon CodeWhisperer.

\textbf{Study setup and Data Collection}: 
The user study included a two-hour morning session, a one-hour lunch break, and a two-hour afternoon session. Participants were introduced to LLMs and CodeWhisperer, followed by hands-on Python tasks using VSCode. They worked at their own pace, recorded their screens, and submitted their work anonymously. Compensation included \$25 CAD per session. 

\textbf{Python programming tasks}: We selected six real-world Python programming tasks involving basic Python, file manipulation, and operating system problems of varying difficulty. We selected tasks used by Xu et al~\cite{xu2022ide}.
Detailed information about the user study and programming tasks can be found in our online appendix \cite{javahar2025cracking}.

\textbf{Data Analysis}: To analyze screen recordings, we used \textit{open} and \textit{axial} coding, key methods in qualitative analysis~\cite{Strauss2015}. 
The first author began with \textit{open coding} on three initial recordings (first tasks of three participants), using MaxQDA~\cite{MAXQDA} to segment and label interactions.
Each time a new code was identified, recordings were revisited for consistency. Codes were reviewed and refined through discussions with two other authors. This process was repeated for recordings of the second and third tasks, with partial participant overlap. Saturation was reached after the third round. Next, we performed \textit{axial coding}, grouping codes into two categories: \textit{CodeWhisperer Suggestions} (i.e., natural language or code prompts) and \textit{Participant Interactions} (i.e., common user behaviors).
Finally, the comprehensive codebook is available in our online appendix~\cite{javahar2025cracking}. 

\subsection{\textcolor{black}{The CodeWatcher Plugin}} 

We developed CodeWatcher~\cite{basha2025codewatcher}, a VSCode extension built with JavaScript and Node.js, to run alongside CodeWhisperer without impacting its functionality. It logs real-time keystrokes and document changes in JSON format. The plugin tracks eight distinct types of user interactions within the IDE during each coding session, on a per-file basis: Start (when editing begins), End (when editing stops), Insertion (adding text), Deletion (removing text), Focus (when a file becomes active), Unfocus (when a file becomes inactive), Copy (copying text), and Paste (pasting text). For each event, the CodeWatcher captures some of the following four \underline{event properties}: (1) Type, (2) Time, (3) Text and (4) Line. Type records which of the eight events was captured. Time represents the timestamp of when there was an increase in the document text larger than three characters at a time. Text corresponds to the changed text if applicable. Line is the text of the line where the change occurred if applicable~\cite{basha2025codewatcher}.

\subsection{Phase 2: Second User Study}
The recruitment strategy for this phase is the same as that of phase one. In this case, the participants included five self-identified men, four self-identified women, and one individual who chose not to disclose their gender identity. All participants were Computer Science students at our university, consisting of one undergraduate and nine graduate students.

\textbf{Study setup}: The setup for this user study mirrored phase one, except participants used our CodeWatcher plugin instead of screen recording. Participants anonymously submitted their completed code files and the associated logs after finishing the tasks to a cloud service. Each participant also received a remuneration of CAD \$25 at the start of each session. 

\subsection{Phase 3: Post Processing and Data Analysis}

During post-processing, we quantitatively analyzed the user event logs to identify and categorize the users' interactions with CodeWhisperer.
Specifically, to answer RQ1, we defined and computed the user interaction variables from the event logs as reported in the Table
\ref{tab:user_interaction_variables}. 
To address RQ2, we analyzed how much of CodeWhisperer’s suggested code was retained by participants. This involved examining \textit{Insertion}, \textit{Copy}, and \textit{Paste} events from CodeWatcher, since only inserted or pasted text could appear in the final code. We categorized events, filtered out invalid or duplicate insertions based on context, and matched valid inserted lines against each participant’s final submitted Python file. A detailed description of the line-matching algorithm is provided separately.

\begin{table*}
\caption{User Interaction Variables with Derivation Approach}
\resizebox{\textwidth}{!}{%
\begin{tabular}{p{0.6cm}p{2.5cm}p{6 cm}p{6.3cm}}
\textbf{} & \textbf{Category} & \textbf{Description} & \textbf{Derivation Approach} \\  \hline \hline
CGI & \textbf{Complete generated insertion} & CodeWhisperer offers a suggestion before the participant begins writing, and the participant accepts the suggestion in its entirety. & Filter for \textit{insertion} objects likely to contain text generated by CodeWhisperer, ensure that \textit{line} contained no additional text preceding \textit{insertion}\\ \hline
PGI & \textbf{Partially generated insertion} & CodeWhisperer offers a suggestion after the participant begins writing, and the participant accepts the suggestion to finish what they have already written. & Filter for \textit{Insertion} objects likely to contain text generated by CodeWhisperer, ensure that \textit{line} contained additional text preceding \textit{insertion}\\ \hline
CSLD & \textbf{Consecutive single letter deletion} & The participant uses the backspace key to erase a word, letter by letter. If multiple single-letter deletions occur within the same second or within a one-second difference, it is counted as a CSLD. & Group \textit{deletion} objects where text is one letter with timestamp difference $<=$ 1 second  \\ \hline
PSLD & \textbf{Partial single letter deletion} & The participant uses the backspace key to erase just one letter. & Isolated single letter deletion object\\ \hline
CD & \textbf{Complete deletion} & The participant highlights and deletes the entire code, leaving nothing but a newline, carriage return, or space. & Filter for deletion objects where line property is empty\\ \hline
PD & \textbf{Partial deletion} & The participant highlights and deletes only a portion of the code, leaving text on the line & Filter for deletion objects where line property is not empty \\ \hline
F & \textbf{Focus} & The participant shifted focus to the IDE/ programming task and CodeWhisperer. & \textit{focus} object\\ \hline
UF & \textbf{Unfocus} & The participant shifted focus away from the programming task and CodeWhisperer. & \textit{unfocus} object  \\ \hline
ES & \textbf{External source} & The participant utilizes outside sources to add to their code, without the help of CodeWhisperer. & \textit{unfocus} object followed by a \textit{paste} object \\ \hline
IS & \textbf{Internal source} & The participant copies and pastes code from within the IDE they are writing in. & \textit{copy} object followed by a \textit{paste} and \textit{insertion}, where text property on \textit{insertion} object corresponds to text property on \textit{copy} object \\ \hline
C & \textbf{Copy} & The participant copies a portion of the code internally & \textit{copy} object\\ \hline \hline
\end{tabular}
}
\label{tab:user_interaction_variables}
\end{table*}

\section{Results}
Some participants showed no recorded interactions with CodeWatcher, possibly due to not using CodeWhisperer suggestions or logging failures. Our dataset includes 2,527 interactions in Task 1, 2,323 in Task 2, and 4,159 in Task 3

\begin{table*}[t]
\centering
\begin{minipage}[t]{0.50\textwidth}
\centering
\caption{User Interactions during Task 1}
\label{tab:participants-interactions1}
\resizebox{\textwidth}{!}{%
\begin{tabular}{lllllllllllll} 
\textbf{User} & \textbf{PGI} & \textbf{CGI} & \textbf{CSLD} & \textbf{PSLD} & \textbf{CD} & \textbf{PD} & \textbf{F} & \textbf{UF} & \textbf{ES} & \textbf{IS} & \textbf{C} & \textbf{Sum} \\\hline \hline
1 & 6 & 3 & 12 & 3 & 1 & 1 & 6 & 6 & 3 & 0 & 1 & 42 \\
2 & 150 & 24 & 105 & 40 & 44 & 34 & 12 & 0 & 7 & 1 & 1 & 418 \\
3 & 0 & 0 &  & 0 & 0 & 0 & 0 & 0 & 0 & 0 &  & 0 \\
4 & 15 & 1 & 22 & 5 & 9 & 10 & 11 & 11 & 3 & 0 & 0 & 87 \\
5 & 77 & 45 & 272 & 35 & 59 & 40 & 62 & 49 & 0 & 10 & 10 & 659 \\
6 & 63 & 46 & 117 & 24 & 32 & 14 & 54 & 59 & 13 & 5 & 11 & 438 \\
7 & 84 & 9 & 55 & 10 & 15 & 20 & 21 & 22 & 6 & 0 & 0 & 242 \\
8 & 33 & 14 & 59 & 9 & 31 & 24 & 53 & 51 & 10 & 3 & 5 & 292 \\
9 & 54 & 3 & 43 & 10 & 11 & 3 & 26 & 26 & 3 & 2 & 2 & 183 \\
10 & 15 & 29 & 36 & 12 & 21 & 14 & 16 & 13 & 6 & 0 & 4 & 166 \\ \hline \hline
Sum & 497 & 174 & 721 & 148 & 223 & 160 & 261 & 237 & 51 & 21 & 34 & 2527 \\
Mean & 49.7 & 17.4 & 80.11 & 14.8 & 22.3 & 16 & 26.1 & 23.7 & 5.1 & 2.1 & 3.77 &  \\
Med & 43.5 & 11.5 & 55 & 10 & 18 & 14 & 18.5 & 17.5 & 4.5 & 0.5 & 2 &  \\
StDev & 46.24 & 17.72 & 80.01 & 13.62 & 19.14 & 13.63 & 22.18 & 21.99 & 4.17 & 3.24 & 4.17 & 
\end{tabular}
}
\end{minipage}
\hfill
\begin{minipage}[t]{0.49\textwidth}
\centering
\caption{User Interactions during Task 2}
\label{tab:participants-interactions2}
\resizebox{\textwidth}{!}{%
\begin{tabular}{lllllllllllll} 
\textbf{User} & \textbf{PGI} & \textbf{CGI} & \textbf{CSLD} & \textbf{PSLD} & \textbf{CD} & \textbf{PD} & \textbf{F} & \textbf{UF} & \textbf{ES} & \textbf{IS} & \textbf{C} & \textbf{Sum} \\\hline \hline
1 & 39 & 23 & 164 & 7 & 30 & 27 & 20 & 20 & 0 & 1 & 4 & 335 \\
2 & 238 & 44 & 116 & 42 & 98 & 47 & 27 & 16 & 11 & 0 & 3 & 642 \\
3 & 28 & 20 & 40 & 4 & 49 & 16 & 18 & 11 & 8 & 0 & 1 & 195 \\
4 & 4 & 0 & 0 & 0 & 0 & 0 & 0 & 0 & 0 & 0 & 0 & 4 \\
5 & 0 & 0 & 0 & 0 & 0 & 0 & 0 & 0 & 0 & 0 & 0 & 0 \\
6 & 110 & 19 & 22 & 13 & 38 & 53 & 77 & 81 & 12 & 1 & 4 & 430 \\
7 & 51 & 10 & 47 & 7 & 8 & 14 & 19 & 13 & 3 & 0 & 0 & 172 \\
8 & 3 & 4 & 17 & 3 & 15 & 18 & 38 & 41 & 10 & 2 & 7 & 158 \\
9 & 66 & 4 & 96 & 10 & 14 & 14 & 38 & 55 & 8 & 0 & 1 & 306 \\
10 & 23 & 6 & 19 & 0 & 10 & 2 & 3 & 3 & 5 & 0 & 0 & 71 \\ \hline\hline
Sum & 562 & 130 & 521 & 86 & 262 & 191 & 240 & 240 & 57 & 4 & 20 & 2313 \\
Mean & 62.44 & 14.44 & 65.12 & 9.55 & 29.11 & 21.22 & 26.66 & 26.66 & 6.33 & 0.44 & 2.5 &  \\
Med & 39 & 10 & 43.5 & 7 & 15 & 16 & 20 & 16 & 8 & 0 & 2 &  \\
StDev & 73.71 & 13.76 & 54.22 & 12.91 & 30.22 & 18.26 & 23.02 & 26.97 & 4.55 & 0.72 & 2.44 & 
\end{tabular}
}
\end{minipage}

\vspace{0.5cm}

\begin{minipage}[t]{0.49\textwidth}
\centering
\captionof{table}{User Interactions  Task 3}
\label{tab:participants-interactions3}
\tiny
\resizebox{\textwidth}{!}{
\begin{tabular}{lllllllllllll} 
\textbf{User} & \textbf{PGI} & \textbf{CGI} & \textbf{CSLD} & \textbf{PSLD} & \textbf{CD} & \textbf{PD} & \textbf{F} & \textbf{UF} & \textbf{ES} & \textbf{IS} & \textbf{C} & \textbf{Sum} \\\hline \hline
1 & 304 & 20 & 312 & 11 & 66 & 160 & 16 & 10 & 9 & 8 & 15 & 931 \\
2 & 0 & 0 &  & 0 & 0 & 0 & 0 & 0 & 0 & 0 &  & 0 \\
3 & 28 & 20 & 40 & 4 & 49 & 16 & 18 & 11 & 8 & 0 & 1 & 195 \\
4 & 95 & 7 & 92 & 24 & 60 & 80 & 13 & 9 & 4 & 18 & 24 & 426 \\
5 & 289 & 53 & 180 & 74 & 104 & 74 & 11 & 6 & 5 & 10 & 11 & 817 \\ 
6 & 155 & 15 & 125 & 35 & 51 & 50 & 53 & 60 & 12 & 3 & 7 & 566 \\
7 & 54 & 7 & 39 & 25 & 14 & 35 & 5 & 5 & 1 & 0 & 0 & 185 \\
8 & 76 & 6 & 20 & 7 & 15 & 25 & 95 & 101 & 16 & 3 & 9 & 373 \\
9 & 106 & 14 & 82 & 38 & 29 & 58 & 74 & 90 & 15 & 6 & 6 & 518 \\
10 & 15 & 17 & 57 & 10 & 16 & 17 & 3 & 3 & 9 & 0 & 1 & 148 \\ \hline\hline
Sum & 1122 & 159 & 947 & 228 & 404 & 515 & 288 & 295 & 79 & 48 & 74 & 4159 \\
Mean & 112.2 & 15.9 & 105.22 & 22.8 & 40.4 & 51.5 & 28.8 & 29.5 & 7.9 & 4.8 & 8.22 &  \\
Med & 85.5 & 14.5 & 82 & 17.5 & 39 & 42.5 & 14.5 & 9.5 & 8.5 & 3 & 7 &  \\
StDev & 107.51 & 14.62 & 91.97 & 22.19 & 31.57 & 46.20 & 33.19 & 38.83 & 5.46 & 5.88 & 7.75 & 
\end{tabular}}
\end{minipage}
\hfill
\begin{minipage}[t]{0.49\textwidth}
\centering
\caption{Percentage and sum of matched lines for Tasks 1, 2, and 3. NLC means Natural Language Comments}
\label{tab:trust}
\tiny
\resizebox{\textwidth}{!}{
\begin{tabular}{lccc|ccc|cccc}
\textbf{} & \multicolumn{3}{c}{\textbf{Tasks 1}} & \multicolumn{3}{c}{\textbf{Tasks 2}} & \multicolumn{3}{c}{\textbf{Tasks 3}} \\ \hline \hline
\textbf{} & \textbf{NLC} & \textbf{Code} & \textbf{Sum} & \textbf{NLC} & \textbf{Code} & \textbf{Sum} & \textbf{NLC} & \textbf{Code} & \textbf{Sum} \\
Generated Insertion Line & 89 & 613 & 702 & 71 & 452 & 523 & 100 & 487 & 587 \\
Lines with match found & 34 & 237 & 271 & 41 & 260 & 301 & 88 & 314 & 402 \\
Altered/Deleted & 55 & 376 & 431 & 30 & 192 & 222 & 12 & 173 & 185 \\
\textbf{Percentage of matches} & 38\% & 39\% & - & 58\% & 58\% & - & 88\% & 65\% & - \\  \hline \hline
\end{tabular}%
}
\end{minipage}
\end{table*}

\subsection{RQ1: How do users interact with and edit code using CodeWhisperer to solve programming tasks, and what detailed interaction patterns emerged from that?}

Table~\ref{tab:participants-interactions1} shows the user interactions recorded for Task 1 (Basic Python programming tasks). These interactions represent the aggregated user interactions for Task 1.1 and Task 1.2. 
In these tasks, consecutive single letter deletions (CSLD) were the most frequent interactions, accounting for 29\% of all actions, followed by partial generated insertions (PGI) at 20\%. Additionally, Task 1 exhibited relatively high rates of focus (F) and unfocus (UF) interactions, constituting 10\% and 9\% of interactions, respectively, which were notably higher compared to other types of interactions. 

For Task 2 (file manipulation programming tasks), the most frequent interactions were PGI (24\%) and CSLD (23\%), maintaining a similar trend to Task 1. Complete generated insertions (CGI) and complete deletions (CD) were also prevalent, comprising 7\% and 11\% of all interactions, respectively. Focus and unfocus interactions remained consistent with Task 1, both at 10\%. Table~\ref{tab:participants-interactions2} shows the user interactions recorded for Task 2. These interactions represent the aggregated user interactions for Task 2.1 and Task 2.2.  In Task 3 (OS programming tasks), PGI remained the most common interaction type at 27\%, followed closely by CSLD at 23\%. Unfocus interactions (UF) were recorded at 7\%, reflecting a moderate level of task engagement within the IDE. Table~\ref{tab:participants-interactions3} shows the user interactions recorded for Task 3. These interactions represent the aggregated user interactions for Task 3.1 and Task 3.2.

\textcolor{black}{Meanwhile, the rates of CGI (Completed Generated Insertions) decrease as the task difficulty increases: percentages go from 6.88\% (Task 1) to 5.62\% (Task 2) to 3.82\% (Task 3). 
This could be explained by the limitations of LLMs in handling more complex programming problems~\cite{Dou2024}.}

To examine interaction patterns, we plotted user interaction variables for each task over normalized time, from task start to completion. The x-axis represents normalized time, the y-axis participant IDs, and a legend distinguishes interaction types. The plots are available in the online appendix. In the coding task T1.2, several participants had long gaps between interactions, possibly reflecting a learning curve, uncertainty in using CodeWhisperer, or reliance on personal knowledge for a simpler task. In contrast, the more challenging T3.1 showed denser interaction patterns, likely due to its complexity prompting more frequent tool use. Focus/unfocus events revealed differences in coding style: some participants worked continuously in the IDE, while others switched contexts more often. Overall, higher task difficulty correlated with increased interactions, suggesting both greater reliance on CodeWhisperer and growing familiarity with its ability to reduce cognitive load.

\textbf{1. Incremental code refinement:} \textcolor{black}{partial} and complete insertions along with consecutive single letter deletions are the most predominant user interaction variables in the Figures. In other words, it was a common occurrence that participants tended to accept end-of-line suggestions, such as completing a comment or finishing the syntax of a function definition then refining portions of it to their liking. 

\textbf{2. Explicit Instruction Using Natural Language Comments:} Participants typically include Natural Language Comments (NLC) containing command words like {\footnotesize\texttt{``Create,"}} {\footnotesize\texttt{``Write,"}} and {\footnotesize\texttt{``Make,"}} 
to prompt the CodeWhisperer and guide it on what to do. 
There exists a clear distinction between participants' comments made for themselves and comments made to interact with CodeWhisperer, the former being formatted in an explanatory manner, detailing how their code works, while the latter utilizing action words for commands and requests. For example, these are some of the comments found in Participant 6's final Python file for Task 3.2:

\begin{lstlisting}[breaklines=true,basicstyle=\footnotesize]
# In the process directory or file names containing dates are renamed such that the date is reformatted from yyyy-mm-dd format to dd-mm-yyyy format.
\end{lstlisting}

Comparatively, these are some of the comments found in the log file that did not make it to the output script but allowed the user to interact with CodeWhisperer:

\begin{lstlisting}[basicstyle=\footnotesize]
 # get the current working directory
 [...]
 # get the date from the file name
\end{lstlisting}

\textbf{3. Base-line structuring with model suggestions:} We also observed how participants tend to accept suggestions that they may not particularly need but are syntactically similar to what they were intending to write. For example, participant 10 used the CodeWhisperer's suggestion to write the phrase {\footnotesize\texttt{``Write a program that prints your favorite color.''}} Using a complete generated insertion, the participant accepted the suggestion in its entirety, only to delete more than half of it until only {\footnotesize\texttt{``Write a program tha"}} was left. This format allowed them to finish the comment with the command they actually wanted, which was {\footnotesize\texttt{``Write a program that also in the process directory or file names containing dates are renamed such that the date is reformatted from yyyy-mm-dd format to dd-mm-yyyy format."}}

\textbf{4. Integrative use with external sources:} In Participant 9's interaction progression (Figure~\ref{fig:task3}), we see a slew of focus and unfocus interactions. This can mean two things: the user decided to (1) exit the IDE 
to consult outside sources when the aid of CodeWhisperer was not sufficient, or (2) take a break, read through the code they have so far, and think about how they can proceed, instead of constantly accepting the end-of-line suggestions offered by CodeWhisperer. 
In Figure~\ref{fig:task1}, for Task 1, this is a lot more prevalent.


\subsection{RQ2: What fraction of CodeWhisperer’s suggestions are retained by users?}

Table~\ref{tab:trust} shows the percentage of CodeWhisperer's suggested NLC and Code that appeared in the final output files submitted by participants during Phase 3 of the second user study. Table~\ref{tab:trust} shows a trend of increasing accepted suggestions being incorporated into the final output code over time. In Task 1 we observe a 38\% match rate for NLC, indicating that over one-third of CodeWhisperer's suggested comments were directly incorporated without edits. This could suggest that the participants made use of the NL comments as explicit instructions to prompt CodeWhisperer on code to write, then deleted these after the code was given. Furthermore, the retention rate for \textit{code} lines in Task 1 is 39\%. 

In Task 2, the match rate for NLC increases significantly to 58\%, representing a 52.6\% increase compared to Task 1. The match rate for code lines also increased to 58\%, reflecting a higher level of reliance on CodeWhisperer's generated code. Again, this might indicate that study participants were getting more comfortable with the tool and its suggestions as they progressed towards the difficult tasks. In Task 3,
the match rate for NLC peaks at 88\%, while the match rate for code lines also rises to 66\%. These results might indicate that by Task 3, participants had developed a significant level of confidence in CodeWhisperer's ability to generate code, as reflected by the high proportion of unedited suggestions included in their final submissions.

\section{Discussion}


\textbf{Behavioral patterns:} 
Users often find significant value in the convenience and efficiency of AI-assisted coding~\cite{groundedCopilot, basha2025trust, Cui2024, novicesUseLLMs, designingTrust, Mendes2024}. 
Aligned with our findings for RQ1, participants often engaged in an iterative process of accepting and refining CodeWhisperer suggestions, indicating they found value in them. This mirrors prior work ~\cite{Mendes2024,vaithilingam2022expectation}, where users viewed AI-generated code, even when incorrect, as useful starting points for crafting correct solutions.

\textcolor{black}{
A similar approach is observed in the context of comments: users accepted suggestions for comments that were syntactically similar to what they were intending to write.} We also observed an integrative use of external sources. That is, there were instances where participants relied on external resources (e.g., Stack Overflow, ChatGPT) when the suggestions provided by CodeWhisperer were perceived as insufficient. Previous studies have shown that users can experience frustration when interacting with code generation tools~\cite{liang2024large, Mendes2024,vaithilingam2022expectation}, which may explain a user's choice to consult an External Source (see Table~\ref{tab:user_interaction_variables}).

\textbf{LLM's suggested code retention:}
Our key finding for RQ2 shows that the proportion of AI-generated content retained by users increased over time. Comment retention rose from 38\% to 88\%, and code retention from 39\% to 65\%. This may be due to two factors: the increasing difficulty of tasks encouraging participants to accept more suggestions, and growing familiarity with the tool, leading to greater trust in its output. Moreover, 
the reliance on CodeWhisperer might allow participants to offload routine work and focus on higher-level problem solving, using the tool to reduce cognitive load.

\textbf{Implications for Tool Builders}
Our four observed patterns: (1) incremental code refinement, (2) use of natural language comments, (3) baseline structuring with suggestions, and (4) integrative use of external sources; offer key insights for improving LLM-based coding tools.

The first and third patterns align with prior work~\cite{groundedCopilot, Mendes2024}, highlighting that both code and comment suggestions serve as valuable starting points. Tools could better support this by allowing users to incrementally accept suggestions, enabling easier customization, a feature that may require IDE redesign~\cite{Wang2024}. In the second pattern, participants used comments for both self-documentation and prompting the LLM. Capturing these guiding comments could help preserve the design rationale behind code generation. The fourth pattern emphasizes the need for external validation. As Kabir et al.~\cite{Kabir2024} found, users overlooked inaccuracies in ChatGPT’s responses 39\% of the time. Tool builders should consider integrating LLMs with trusted search or documentation sources to offer more accurate, context-aware support when AI alone is insufficient.






\textbf{Threats to Validity:} Our small sample limits our ability to distinguish the effects of demographics versus individual coding styles, so larger cohorts are needed for generalizability. CodeWatcher logs document changes and keystrokes, then infers AI-generated code.  This post-processing, combined with our small sample, risks invalid or outlier observations. Finally, both participants familiar with the tool and those unaware of its functionality could skew results (e.g., missing interactions in Figures ~\ref {fig:task1} and ~\ref{fig:task3} may reflect logging failures).

\section{Conclusion and Future Work}
This paper examines user interactions with LLM-based code generators, focusing on behavior, trust, and perceptions during programming tasks. We investigated how users engage with these tools, how their trust evolves, and how they perceive benefits in productivity, efficiency, and learning.

Future work should explore the long-term impact of LLM use on skill development, particularly the potential erosion of foundational coding skills among novices. Studying usage patterns and trust across diverse demographic groups with larger samples could also offer valuable insights. Understanding these long-term effects and optimizing tool design for both learning and productivity remains a key research priority.

\section*{Acknowledgment}
We thank GOOGLECA (PWPU GR028067) and CNPq (420406/2023-9 and 442779/2023-2) for funding this research.


\balance
\bibliographystyle{acm}
\bibliography{ref}

\newpage
\appendix
\begin{table}[h!]
\centering
\caption{CodeBook of Events}
\begin{tabular}{p{4cm} p{11cm}}
\hline
\textbf{Event} & \textbf{Description} \\
\hline
\multicolumn{2}{l}{\textbf{CodeWhisperer Events}} \\
PromptSuggestion & CodeWhisperer makes a suggestion to complete the comment the user is currently typing \\
CodeSuggestion & CodeWhisperer suggests a code block \\
PromptCodeSuggestion & CodeWhisperer makes a suggestion to complete the comment the user is currently typing and a corresponding code block \\
\hline
\multicolumn{2}{l}{\textbf{Participant Events}} \\
PromptCreation & Participant begins writing a prompt \\
PromptCompletion & Participant completes the prompt themselves \\
PromptSuggestionAcceptExplicit & Participant accepts a CodeWhisperer prompt suggestion in its entirety \\
PromptSuggestionAcceptImplicit & Participant manually types prompt similar to CodeWhisperer's suggestion \\
PromptEdit\textsuperscript{\ref{fn:edit},\ref{fn:partialdel}} & Participant edits a prompt \\
PromptSuggestionEdit & Participant edits an accepted prompt \\
PromptSuggestionDelete & Participant deletes an accepted prompt suggestion \\
PromptDeletion & Participant deletes a prompt in its entirety \\
CodeCreation & Participant begins writing code \\
CodeCompletion & Participant completes a stream of code themselves \\
CodeDeletion & Participant deletes a stream of code they wrote themselves \\
CodeSuggestionAcceptExplicit & Participant accepts a CodeWhisperer code suggestion in its entirety \\
CodeSuggestionAcceptImplicit & Participant manually types code similar to CodeWhisperer's suggestion \\
CodeEdit & Participant edits a code section untouched by generated code \\
CodeSuggestionEdit\textsuperscript{\ref{fn:edit}} & Participant edits an accepted code suggestion \\
CodeSuggestionDelete & Participant deletes an accepted code suggestion \\
CodeSuggestionPartialDelete & Participant deletes parts of a generated code block \\
SuggestionInduction & CodeWhisperer fails to make a suggestion, making the participant go back and forth to induce a suggestion \\
ProgramRunSuccessful & The participant runs the program successfully, with expected output \\
ProgramRunUnsuccessful & The participant runs the program unsuccessfully, with unexpected output \\
PromptSuggestionExplore & The participant cycles through the prompt suggestions \\
CodeSuggestionExplore & The participant cycles through code suggestions \\
UnusualActivity & Participant performs an action prohibited in the experiment (e.g., using other code-generation tools, pasting from the repository) \\
\hline
\end{tabular}

\vspace{0.5em}
\raggedright
\textsuperscript{\ref{fn:edit}}\label{fn:edit} An edit is considered continuous alteration until the cursor moves away from the section being edited. \\
\textsuperscript{\ref{fn:partialdel}}\label{fn:partialdel} Partial deletion is considered an edit for prompts.
\end{table}
\newpage

\begin{table}[h]
\centering
\begin{tabular}{l p{8cm} p{8cm}}
\hline
\textbf{Task} & \textbf{Description} & \textbf{Example Output} \\ \hline
T1-1 & After running \texttt{python3 T1-1.py}, the program randomly generates 100 lower-cased characters (a--z) and 100 integers (1--20 inclusive). These are paired into a dictionary where the key is the character and the value is the sorted list of integers associated with it. The dictionary is printed one key per line, sorted alphabetically. & \texttt{\$ python3 T1-1.py} \newline \texttt{a 3 5 6} \newline \texttt{b 1 1 2 4 5} \newline \texttt{c 19 ...} \newline \texttt{z 1 12 20} \\ \hline
T1-2 & After running \texttt{python3 T1-2.py}, the program prints the date and time one week from now in GMT timezone in \texttt{mm-dd-yyyy hh:mm} (24hr) format. & \texttt{\$ python3 T1-2.py} \newline \texttt{04-23-2020 16:33} \\ \hline
T2-1 & After running \texttt{python3 T2-1.py}, the program deletes the first and last columns from \texttt{data.csv} and saves the result to \texttt{output/output.csv}. & \texttt{\$ python3 main.py} \newline \texttt{\$ less output/output.csv} \newline \texttt{first\_name,last\_name,email,gender} \newline \texttt{Merry,MacKettrick,mmackettrick0@latimes.com,Male} \newline \texttt{Calla,Truin,ctruin1@surveymonkey.com,Female} \\ \hline
T2-2 & After running \texttt{python3 T2-2.py}, trims leading/trailing whitespaces and blank lines for all text files in \texttt{data/}, normalizes newlines to LF, and converts ISO-8859-15 text files to UTF-8. Non-text files are unchanged. Output saved to \texttt{output/}. & \texttt{\$ python3 main.py} \newline \texttt{\$ ls output} \newline \texttt{aaa.txt bbb.txt ccc.txt ddd.png} \newline \texttt{\$ file bbb.txt} \newline \texttt{UTF-8 Unicode text} \\ \hline
T3-1 & After running \texttt{python3 T3-1.py}, copies all files/dirs under \texttt{data/} to \texttt{output/}, renaming any dates in filenames from \texttt{yyyy-mm-dd} to \texttt{dd-mm-yyyy} format. & \texttt{\$ python3 T3-1.py} \newline \texttt{\$ ls data} \newline \texttt{Photos\_2019-03-22 ccc.txt data-02-01-1994.txt 'mybook at 2020-04-01.txt'} \newline \texttt{\$ ls output} \newline \texttt{Photos\_22-03-2019 ccc.txt data-02-01-1994.txt 'mybook at 01-04-2020.txt'} \\ \hline
T3-2 & After running \texttt{python3 T3-2.py}, processes subdirectories of \texttt{data/}: \newline -- For dirs with \texttt{.txt} files: concatenate them (alphanumeric order) into one \texttt{.txt} file named after the dir. \newline -- For dirs with \texttt{.json} files: merge lists, re-number \texttt{id} fields starting from 1, save as \texttt{.json}. \newline -- Ignore other dirs. & \texttt{\$ python3 T3-2.py} \newline \texttt{\$ ls data} \newline \texttt{filelist roster todo} \newline \texttt{\$ ls output} \newline \texttt{filelist.txt roster.json} \\ \hline
\end{tabular}
\caption{Task Descriptions and Example Outputs}
\label{tab:tasks}
\end{table}

\newpage
\section{Additional Figures}


\begin{figure}[ht]
    \centering
    \begin{subfigure}[b]{\textwidth}
        \centering
        \includegraphics[width=0.7\linewidth]{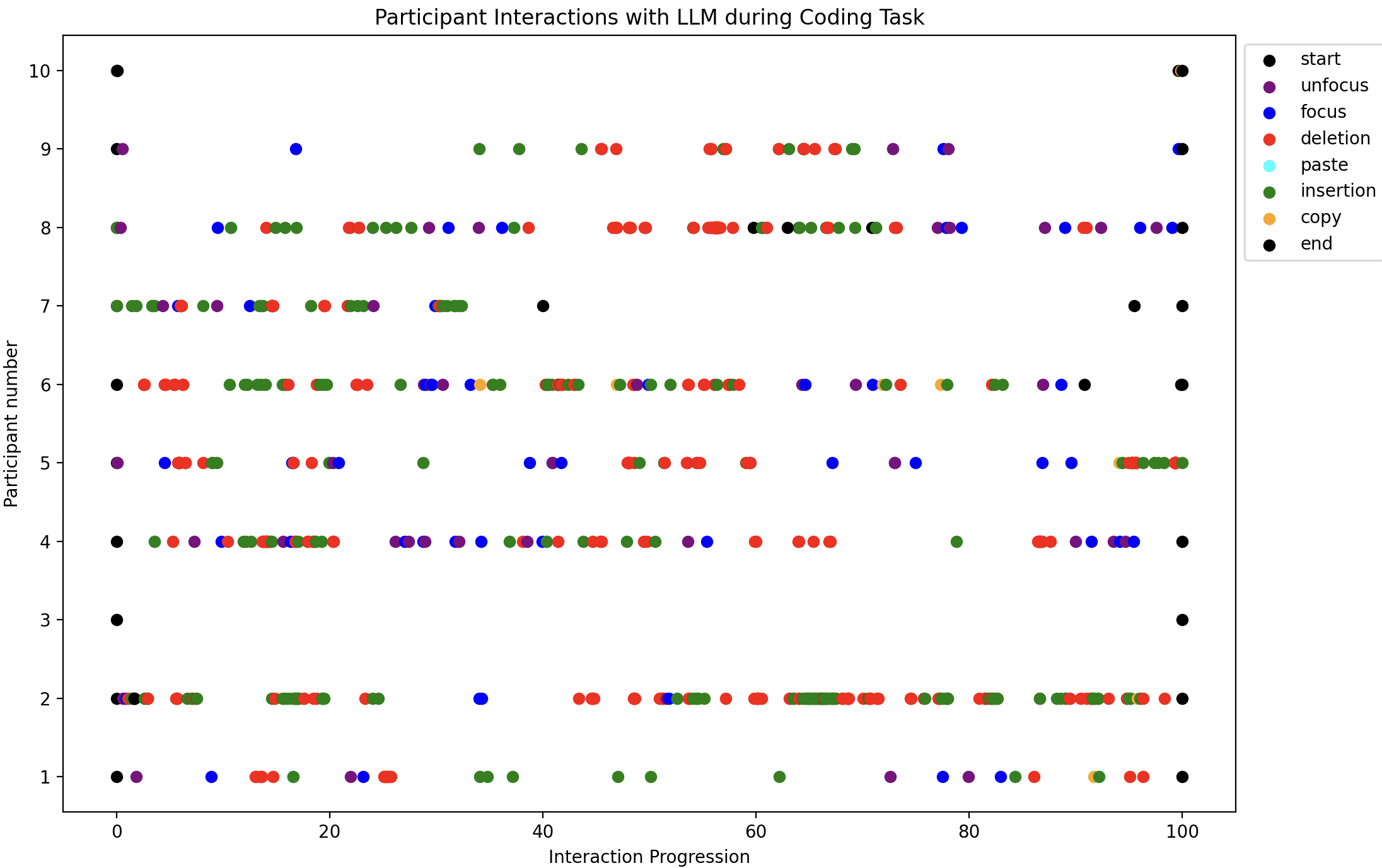}
        \caption{Participant Interactions during Coding Task 1.1}
        \label{fig:task1}
    \end{subfigure}
    
    \vspace{1em} 

    \begin{subfigure}[b]{\textwidth}
        \centering
        \includegraphics[width=0.7\linewidth]{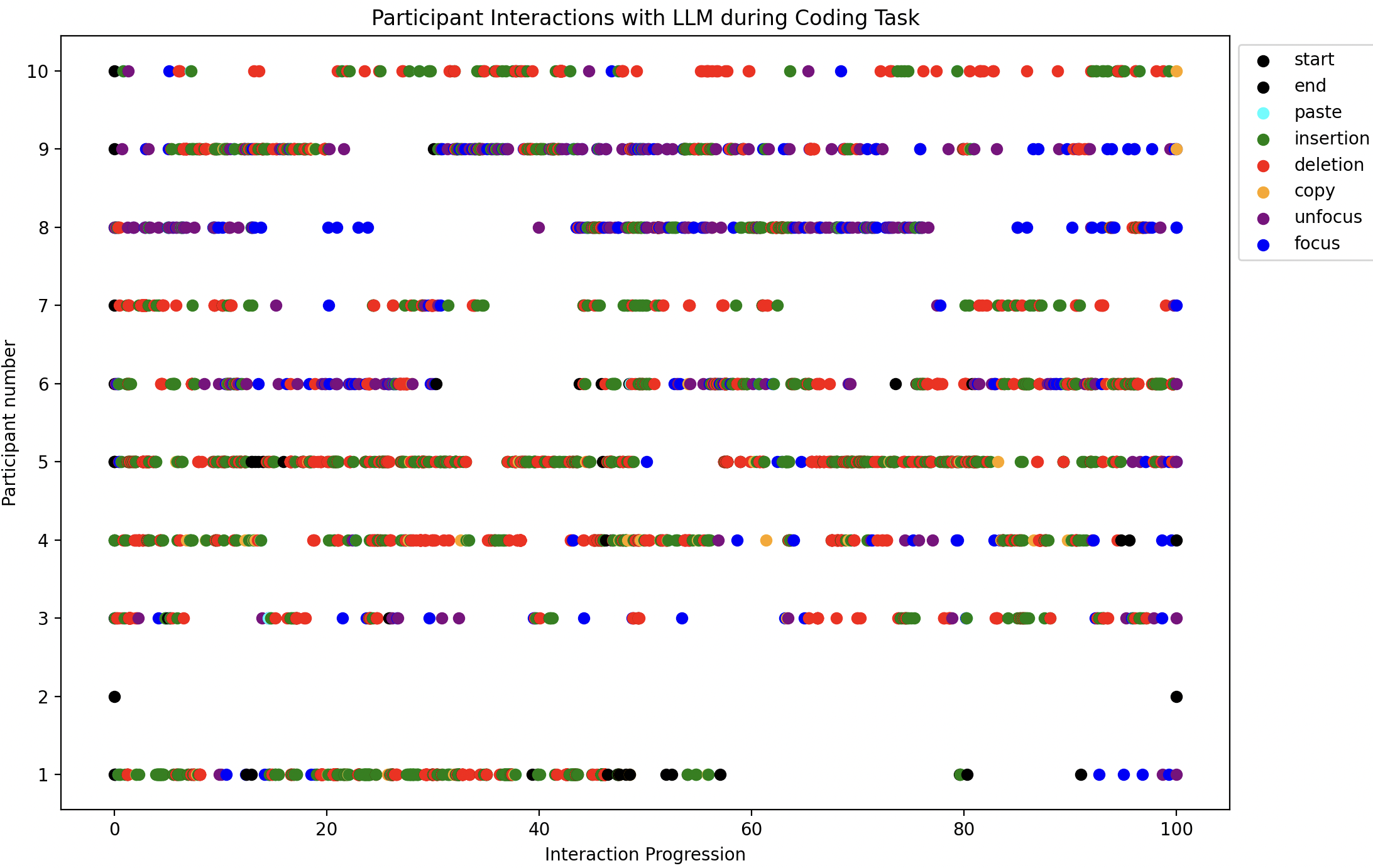}
        \caption{Participant Interactions during Coding Task 3.1}
        \label{fig:task3}
    \end{subfigure}
    
    \label{fig:task1-3-combined}
    \caption{Comparison of participant interactions across two tasks. To simplify the Figure, we aggregated related interaction variables. For instance, CGI and PGI are presented only as ``insertions''.}
\end{figure}

\end{document}